\documentclass[onecolumn,journal,twoside]{IEEEtran}

\usepackage[T1]{fontenc}    
\usepackage{hyperref}       
\usepackage{url}            
\usepackage{booktabs}       
\usepackage{amsfonts}       
\usepackage{nicefrac}       
\usepackage{microtype}      
\usepackage[lofdepth,lotdepth]{subfig}
\usepackage[pdftex]{graphicx}
\usepackage{amsmath,amsthm,amssymb}
\usepackage{enumerate}
\usepackage{enumitem}
\usepackage{cite}
\usepackage{bm}
\usepackage{bbm}
\usepackage{cases}
\usepackage[countmax]{subfloat}
\usepackage{multirow}
\usepackage{verbatim}

\usepackage{pgfplots}
\usepackage{tcolorbox}

\newtheorem{theorem}{Theorem}

\theoremstyle{definition}

\newtheorem{definition}{Definition}

\theoremstyle{remark}
\newtheorem{remark}{Remark}

\newcommand{\vct}[1]{\bm{#1}}

\title{Polynomially Coded Regression: Optimal Straggler Mitigation via Data Encoding\\
}

\author{Songze~Li, Seyed~Mohammadreza~Mousavi~Kalan, Qian~Yu, Mahdi~Soltanolkotabi, and A.~Salman~Avestimehr\\
Department of Electrical Engineering, University of Southern California, Los Angeles, CA, USA
}

\begin{document}
\sloppy

\maketitle

\begin{abstract}
    We consider the problem of training a least-squares regression model on a large dataset using gradient descent. The computation is carried out on a distributed system consisting of a master node and multiple worker nodes. Such distributed systems are significantly slowed down due to the presence of slow-running machines (stragglers) as well as various communication bottlenecks. We propose ``polynomially coded regression'' (PCR) that substantially reduces the effect of stragglers and lessens the communication burden in such systems. The key idea of PCR is to encode the partial data stored at each worker, such that the computations at the workers can be viewed as evaluating a polynomial at distinct points. This allows the master to compute the final gradient by interpolating this polynomial. PCR significantly reduces the recovery threshold, defined as the number of workers the master has to wait for prior to computing the gradient. In particular, PCR requires a recovery threshold that scales inversely proportionally with the amount of computation/storage available at each worker. In comparison, state-of-the-art straggler-mitigation schemes require a much higher recovery threshold that only decreases linearly in the per worker computation/storage load. We prove that PCR's recovery threshold is near minimal and within a factor two of the best possible scheme.  Our experiments over Amazon EC2 demonstrate that compared with state-of-the-art schemes, PCR improves the run-time by $1.50\times \!\sim\! 2.36\times$ with naturally occurring stragglers, and by as much as $2.58\times \!\sim\! 4.29\times$ with artificial stragglers.
\end{abstract}
\section{Introduction}

Modern machine learning models have achieved unprecedented performance on a wide range of complex tasks, such as object detection and face recognition. These models often consist of hundreds or even thousands of layers involving hundreds of millions of parameters, which need to be trained over massively large datasets. As a result, many distributed learning systems have been developed to parallelize the storage and processing of the data onto multiple cores on a single machine, or multiple machines in computing clusters (see, e.g.,~\cite{recht2011hogwild,gemulla2011large,zhuang2013fast,seide20141}). 


%

Gradient-based methods such as Gradient Descent (GD) serve as a workhorse for training such models by iteratively refining learning models over the training data. To scale gradient methods to handle massive amounts of training data, developing parallel/distributed implementations of gradient descent over multiple cores or GPUs on a single machine, or multiple machines in computing clusters is crucial. A common approach to distributing GD is via a master/worker system where the training data is distributed by a master node across multiple worker nodes. Each worker computes a partial gradient based on its locally stored data partition and sends it to the master, who aggregates all partial gradients to update the model parameters.

However, as we scale out training and computations across many distributed nodes new challenges arise. For instance, in the master/worker system mentioned above, the master needs to wait for the results from all workers to aggregate the full gradient. Therefore, the run-time of each iteration of distributed GD may be limited by the slowest worker (in terms of computation or communication). These slow or \emph{stragglar} nodes significantly slow down distributed GD and have been widely observed to be a major performance bottleneck in distributed computing systems in general~\cite{zaharia2008improving,ananthanarayanan2013effective,dean2013tail}. For example, it was experimentally demonstrated in~\cite{zaharia2008improving} that the straggler's effect can prolong the job execution time by as much as 5 times. Furthermore, as we distribute computations across many nodes, massive amounts of partially computed data must be moved between them. 
This is frequently performed over many iterations 
and creates a substantial \emph{communication} bottleneck (e.g., to/from the master node).



There has recently been an exciting surge of coding-theoretic strategies to mitigate straggler and bandwidth bottlenecks (see, e.g.,~\cite{LMA_all,li2016fundamental,lee2015speeding,LMA16_unify,dutta2016short,yu2017polynomial,karakus2017straggler,maityrobust}). The key idea behind these strategies is to inject computation redundancy in coded and/or structured forms to avoid stragglers and reduce the communication load. In particular, in the context of the above distributed GD architecture, ``gradient coding''(GC) has been proposed in~\cite{TLDK-ICML} to effectively leverage extra computation/storage at the the workers in order to enable the master node to tolerate missing results from a random subset of stragglers.  For a system consisting of $n$ workers and a training dataset partitioned into $n$ batches, the core idea underlying GC is to first allocate $1 \leq r \leq n$ data batches onto each worker, and then design a linear combination of the $r$ partial gradients computed from the local data batches, and send it to the master. These linear combinations are designed such that the master can recover the full gradient (i.e., the sum of all partial gradients) by linearly combining the results from \emph{any} subset of $n-r+1$ workers, achieving a robustness to $r-1$ stragglers. In other words, GC achieves a recovery threshold (i.e., the number of workers that the master needs to wait for in order to compute the final gradient), denoted by $K_{\textup{GC}}(r)$, of
\begin{align}
    K_{\textup{GC}}(r)=n-r+1.\label{eq:GCrecover}
\end{align}

There have been several works proposed in the literature to further improve GC~\cite{halbawi2017improving,raviv2017gradient,li2018near,ye2018communication}. In particular, a ``batched coupon's collector''(BCC) algorithm was proposed in~\cite{li2018near}, which utilizes random data placement to achieve an \emph{average} recovery threshold $K^{\textup{average}}_{\textup{BCC}}=\lceil \frac{n}{r} \rceil \log \lceil \frac{n}{r} \rceil$. Also in a very recent work~\cite{ye2018communication}, novel techniques have been developed to code across coordinates of the partial gradients, trading a larger recovery threshold for less amount of communication from each worker. In spite of these recent advances, however, the state-of-the-art minimum recovery threshold for the worst case scenario remains to be (\ref{eq:GCrecover}). Such high recovery thresholds also create a major communication bottleneck as the master needs to wait for many workers to finish and send their results. 

\subsection{Contributions}

We focus on least-squares regression problems, such as those appearing in signal estimation or  kernel-based learning tasks. We consider  distributed gradient decent in master/worker architecture for such problems.
We propose a novel scheme, named ``polynomially coded regression'' (PCR), which achieves near-optimal robustness to stragglers by significantly improving over the state-of-the-art recovery threshold of GC. Specifically, in a system consisting of $n$ workers and a training dataset partitioned into $n$ batches, we prove that PCR achieves a recovery threshold of 
\begin{align}
   K_{\textup{PCR}} = 2\lceil\tfrac{n}{r}\rceil-1, \label{eq:PCRrecover}
\end{align}
where $r$ is the number of data batches that is processed at each worker. Comparing the recovery threshold of PCR (\ref{eq:PCRrecover}) to that of GC (\ref{eq:GCrecover}), we note that PCR enables a recovery threshold that is inversely proportional to the number of batches that is processed at each worker, significantly improving over state-of-the-art.

Furthermore, we show that the recovery threshold for distributed regression under any scheme is lower bounded by $\lceil\frac{n}{r}\rceil$. Hence, PCR achieves the optimal recovery threshold to within a constant factor of 2. 

The key idea of PCR is to \emph{encode data}, rather than the partial gradients computed from the uncoded data as done in the GC schemes. Exploiting the particular algebraic property of gradient computation for least-squares regression problems, the proposed PCR scheme designs coded data batches as linear combinations of the uncoded batches, and stores them on the workers. Then, within each GD iteration, each worker computes its local result by directly operating on the coded data batches, which is effectively evaluating a polynomial of degree $2\lceil \frac{n}{r}\rceil-2$ at a particular point. Finally, the master interpolates the underlying polynomial using the computation results from the fastest $2\lceil \frac{n}{r}\rceil-1$ workers, which is used to find the full gradient via polynomial evaluations.

Compared with the GC schemes~\cite{TLDK-ICML,halbawi2017improving,raviv2017gradient,ye2018communication}, for the same computation/storage overhead $r$, PCR significantly reduces the recovery threshold within each GD iteration (approximately by a multiplicative factor of $r/2$). PCR also has a lower decoding complexity at the master. Surprisingly, PCR achieves these reductions while maintaining the same amount of communication from each worker to the master.  We summarize these comparisons in Table \ref{Tab1}. 


We also experimentally evaluate the performance of PCR scheme on Amazon EC2 clusters. These experiments suggest that compared with GC scheme, PCR improves the overall run-time of distributed GD for least-squares regression by $1.50\times \!\sim\! 2.36\times$ in the presence of naturally occurring stragglers, and by $2.58\times \!\sim\! 4.29\times$ when artificial stragglers were introduced. We provide a sample result from our experiments in Table \ref{Tab2}.


\begin{table}[htbp]
\centering
\subfloat[Subtable 1 list of tables text][Theoretical comparisons.]{
\begin{tabular}{|c|c|c|}
\hline
& \# batches processed & \multirow{2}{*}{recovery threshold}\\
& at each worker & \\
\cline{1-3}
\hline
\rule{0pt}{10pt} GC & $r$ & $n-r+1$ \\
\hline
\rule{0pt}{10pt} PCR & $r$ & $2\lceil\tfrac{n}{r}\rceil-1$ \\
\hline 
\end{tabular}\label{Tab1}}
\qquad 
\subfloat[Subtable 2 list of tables text][Experiments for 100 GD iterations. Each worker was artificially delayed for 0.5 s with probability 5\% in each iteration.]{
\begin{tabular}{|c|c|c|c|}
\hline
& \multirow{2}{*}{\# workers} & \# batches processed & \multirow{2}{*}{run-time}\\
& & at each worker &\\
\cline{1-3}
\hline
\rule{0pt}{10pt} GC & 40 & 10 & 16.821 s \\
\hline
\rule{0pt}{10pt} PCR & 40 & 10 & 3.925 s \\
\hline
\end{tabular}\label{Tab2}}
\vspace{1mm}
\caption{Performance comparison between Gradient Coding (GC) schemes~\cite{TLDK-ICML,halbawi2017improving,raviv2017gradient,ye2018communication} and the proposed Polynomially Coded Regression (PCR) for running distributed least-squares regression on a system consisting of $n$ workers and a training dataset partitioned into $n$ data batches.}
\end{table}

\section{Problem setting and main results}
We now describe the distributed regression problem that we focus on in this paper, define our evaluation metrics for algorithms, and finally state the main theoretical contributions.


\subsection{Distributed gradient descent for regression problems}
We focus on linear regression problems with a least-squares objective. Given a training dataset consisting of $m$ feature inputs $\vct{x}_i\in\mathbb{R}^d$ and labels $y_i\in\mathbb{R}$ we wish to find the coefficients $\vct{w}\in\mathbb{R}^d$ of a linear function $\vct{x}\mapsto\langle \vct{x},\vct{w}\rangle$ that best fits this training data. Minimizing the empirical risk leads to the following optimization problem
\begin{align}
\label{mainopt}
   \underset{\vct{w}\in\mathbb{R}^d}{\min}\text{ } {\cal L}(\vct{w}) = \frac{1}{m} \sum_{i=1}^{m} (\vct{x}_i^\top \vct{w}-y_i)^2 = \frac{1}{m} ||\vct{X} \vct{w} - \vct{y}||^2.
\end{align}
Here, $\vct{X} = [\vct{x}_1 \; \vct{x}_2 \cdots  \;\vct{x}_{m}]^\top \in \mathbb{R}^{m \times d}$ is the feature matrix and $\vct{y}=[y_1 \; y_2 \cdots  \;y_{m}]^\top \in \mathbb{R}^m$ is the output vector obtained by concatenating the input features and output labels, respectively.

We would like to note that many \emph{nonlinear} regression problems can also be written in the form above. In particular, consider the problem of finding the best function $h$ belonging to a hypothesis class $\mathcal{H}$ that fits the training data 
\begin{align}
   \underset{h\in\mathcal{H}}{\min}\text{ }{\cal L}(h) = \frac{1}{m} \sum_{i=1}^{m} (h(\vct{x}_i)-y_i)^2.
\end{align}
Such nonlinear regression problems can often be cast in the form \eqref{mainopt}, and be solved efficiently using the so called kernalization trick \cite{scholkopf2001generalized}. Our data encoding scheme is amenable to the kernalization trick and easily applies to such problems. However, for simplicity of exposition in the remainder of the paper we focus on the simpler instance \eqref{mainopt}.


A popular approach to solving problems like the above is via gradient descent (GD). In particular, GD iteratively refines the weight vector $\vct{w}$ by moving along the negative gradient direction via the following updates
\begin{align}
    \vct{w}^{(t+1)} = \vct{w}^{(t)} - \eta^{(t)} \nabla {\cal L}(\vct{w}^{(t)}) = \vct{w}^{(t)} - \eta^{(t)} \frac{2}{m} \vct{X}^\top(\vct{X}\vct{w}^{(t)}-\vct{y}).\label{eq:GD-update}
\end{align}
Here, $\eta^{(t)}$ is the learning rate in the $t$th iteration.

\begin{figure}[htbp]
  \centering
  \includegraphics[width=0.7\textwidth]{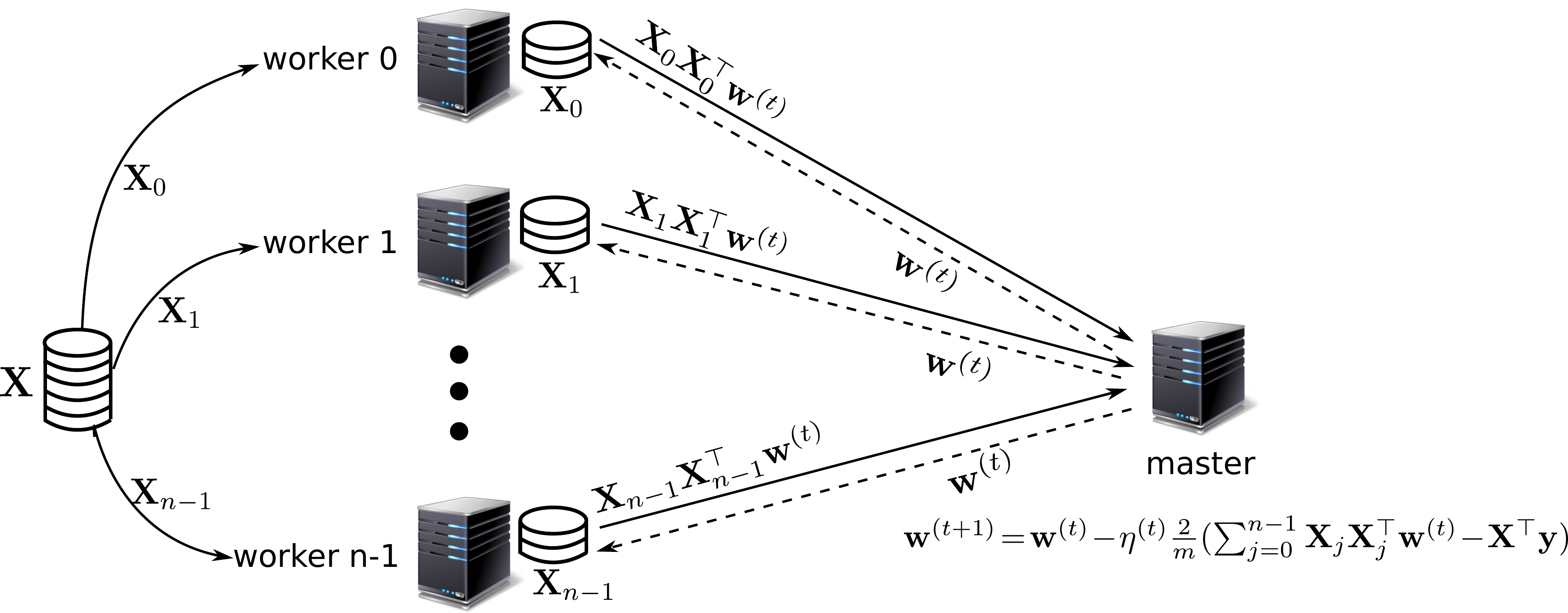}
  \caption{An illustration of a master/worker architecture for data-parallel distributed linear regression.}
  \label{fig:DR}
\end{figure}

When the size of the training data is too large to store/process on a single machine, the GD updates can be calculated in a distributed fashion over many computing nodes. As illustrated in Figure~\ref{fig:DR}, we consider a master/worker architecture that consists of a master node and $n$ worker nodes. Using a naive data-parallel distributed regression scheme, we first partition the input data matrix $\vct{X}$ into $n$ equal-sized sub-matrices such that $\vct{X} = [\vct{X}_0 \; \vct{X}_1 \cdots  \;\vct{X}_{n-1}]^\top$, where each sub-matrix $\vct{X}_j \in \mathbb{R}^{d \times \frac{m}{n}}$ contains $\frac{m}{n}$ input data points, and is stored on worker~$j$. Within each iteration of the GD procedure, the master broadcasts the current weight vector $\vct{w}$ to all the workers. Upon receiving $\vct{w}$, each worker~$j$ computes $\vct{X}_j\vct{X}_j^\top\vct{w}$, and returns it to the master. The master waits for the results from all workers and sums them up to obtain the full gradient
\begin{align}
    \vct{X}^\top\vct{X}\vct{w} = \sum_{j=0}^{n-1} \vct{X}_j \vct{X}_j^\top \vct{w}.\label{eq:linear-grad}
\end{align}
Then, the master uses this gradient to update the weight vector via (\ref{eq:GD-update}). \footnote{Since the value of $\vct{X}^\top \vct{y}$ does not vary across iterations, it only needs to be computed once. We assume that it is available at the master for weight updates.}


\subsection{Coded computation schemes and their recovery thresholds}
\label{sec:probFormulation}
The naive distributed scheme discussed in the previous sub-section requires the master to wait for results from all the workers. Therefore, even a single straggler can significantly delay progress in each iteration. In order to overcome the effect of such stragglers, we can have the workers store and process redundant data such that the computation can be accomplished with the results from only a subset of workers. For example, each worker, instead of 1, stores and processes $1< r \leq n$ sub-matrices. Then, we can partition the $n$ sub-matrices into $\frac{n}{r}$ batches of size $r$, and repeatedly store each batch on $r$ workers. Utilizing this storage/computation redundancy, in the worst case, the master needs the results returned from the fastest $n-r+1$ workers to compute the final gradient. This example can be viewed as applying ``repetition code'' to distributed GD, where we trade $r$ times redundant storage/computation for tolerating $r-1$ stragglers. In general, for a given storage/computation load, we can design optimal coding techniques to minimize the number of workers the master needs to wait for before recovering the gradient. Motivated by this idea, we consider a general distributed regression framework with an input feature matrix $\vct{X} = [\vct{X}_0 \; \vct{X}_1 \cdots  \;\vct{X}_{n-1}]^\top$ and $n$ workers, denoted by $\textup{worker } 0,\ldots,\textup{ worker } n-1$. Each worker~$j$ stores $r$ ($1 \leq r \leq n$) matrices locally, each of which has the same dimension as a sub-matrix $\vct{X}_j$. In each iteration, each worker performs local computation utilizing the received weight vector $\vct{w}$ and the locally stored data. The master waits for the results from a subset ${\cal N} \subseteq \{0,1,\ldots,n-1\}$ of workers, and uses them to recover the desired computation in (\ref{eq:linear-grad}). For this framework, a coded computation scheme consists of the following elements.
\begin{itemize}[leftmargin=*]
     \item \textbf{Computation/storage parameter.} We characterize the computation/storage load at each worker via a parameter $r \in \{1,2,\ldots,n\}$. Specifically, each worker stores some data generated from the feature matrix $\vct{X}$ whose size is $\frac{r}{n}$-fraction of the size of $\vct{X}$.
     \item \textbf{Encoding functions.}
     We encode the data stored at the workers via a set of $n$ encoding functions $\vct{\rho} =( \rho_0,\rho_1,\ldots,\rho_{n-1})$ where $\rho_{j}$ is the encoding function of worker $j$. Each encoding function $\rho_j$ maps the input data $\vct{X}$ into $r$ coded sub-matrices $\tilde{\vct{X}}_{j,0},\tilde{\vct{X}}_{j,1},\ldots, \tilde{\vct{X}}_{j,r-1} \in \mathbb{R}^{d \times \frac{m}{n}}$ which are locally stored at worker~$j$. In particular, each $\tilde{\vct{X}}_{j,k}$, $k=0,1,\ldots,r-1$, is a linear combination of the sub-matrices $\vct{X}_0, \vct{X}_1,\ldots,\vct{X}_{n-1}$, i.e.,
     \begin{align}
         \tilde{\vct{X}}_{j,k} = \sum_{i=0}^{n-1} a_{j,k,i} \vct{X}_i.
     \end{align}
     Here, the coefficients $a_{j,k,i}$ are specified by the encoding function $\rho_j$ of worker $j$.
     \item \textbf{Computation functions.} Each worker uses the $r$ encoded sub-matrices along with the weight vector $\vct{w}$ received from the master to perform its computation. We use $\phi_j: \mathbb{R}^{d \times \frac{m}{n} \times r} \times \mathbb{R}^d \rightarrow \mathbb{R}^{\ell_j}$ to denote this mapping whose output is an arbitrary length-$\ell_j$ vector that is computed locally at worker $j$ using $\tilde{\vct{X}}_{j,0},\tilde{\vct{X}}_{j,1},\ldots, \tilde{\vct{X}}_{j,r-1}$ and $\vct{w}$. Thus we have a set of $n$ computation functions $\vct{\phi} = (\phi_0,\phi_1,\ldots,\phi_{n-1})$ one for each worker.
  
     \item \textbf{Decoding function.} The master uses a decoding function $\psi: \underset{j \in {\cal N}}{\times}\mathbb{R}^{\ell_j} \rightarrow \mathbb{R}^{d}$ to map the computation results of the available workers in ${\cal N}$ to the desired computation $ \vct{X}^\top\vct{X}\vct{w}$.
\end{itemize}

Now that we have discussed the different components of coded computation schemes, we need a metric to compare them in the presence of stragglers.
\begin{definition}
For an integer $k$, we say that a computation scheme $S = (\vct{\rho},\vct{\phi},\psi)$ is \emph{$k$-recoverable} if the master can recover $\vct{X}^\top\vct{X}\vct{w}$ from the local computation results of \emph{any} $k$ out of $n$ workers. We define the recovery threshold of a computation scheme $S$ with a computation/storage load $r$ at each worker, denoted by $K_S(r)$, as the minimum value of $k$ such that $S$ is $k$-recoverable.
\end{definition}
Consider a distributed linear regression task executed on $n$ workers with a local computation/storage load $r$ each. We are interested in finding the minimum recovery threshold achieved among all computation schemes along with the corresponding scheme. This \emph{optimal recovery threshold} can be formally defined as  
\begin{align} \label{eq:ORT}
    K^*(r) := \underset{S}{\min} K_{S}(r).
\end{align}
We would like to note that state-of-the-art gradient coding (GC) schemes (see, e.g.,~\cite{TLDK-ICML,halbawi2017improving,raviv2017gradient,ye2018communication}) achieve a recovery threshold of $K_{\textup{GC}}(r) = n-r+1$. To see this first note that each worker stores $r$ uncoded sub-matrices for local processing. For example, using the cyclic repetition scheme in~\cite{TLDK-ICML}, worker $j$ stores $\vct{X}_{j}, \vct{X}_{j+1}, \ldots, \vct{X}_{j+r-1}$ locally, and sends a liner combination of the computation results $\vct{X}_{j}\vct{X}_{j}^\top\vct{w}, \vct{X}_{j+1}\vct{X}_{j+1}^\top\vct{w}, \ldots, \vct{X}_{j+r-1}\vct{X}_{j+r-1}^\top\vct{w}$ to the master, who can recover the final result $\vct{X}_{0}\vct{X}_{0}^\top\vct{w}+\cdots+\vct{X}_{n-1}\vct{X}_{n-1}^\top\vct{w}$ by linearly combining the messages received from any subsets of $n-r+1$ workers. This leads us to the main challenge of this paper

\begin{tcolorbox}
\begin{itemize}[leftmargin=*]
    \item Is it possible to develop coded schemes with significantly better recovery thresholds?
    \item What is a good lower bound on the recovery threshold of any coded scheme?
\end{itemize}  
\end{tcolorbox}



\subsection{Main result}
We present our main result in the following theorem.
\begin{theorem}\label{thm:newCoding}
Consider the distributed linear regression problem (as defined in Section~\ref{sec:probFormulation}) executed over $n$ workers, each with a computation/storage parameter of $r \in \{2,\ldots,n\}$. Our proposed coded computation scheme called Polynomially Coded Regression (PCR), achieves the optimal recovery threshold to within a factor of $2$. That is 
\begin{align}\label{eq:PCR}
    \tfrac{1}{2}K_{\textup{PCR}}(r) < K^*(r) \leq K_{\textup{PCR}}(r)= 2\lceil \tfrac{n}{r}\rceil -1,
\end{align}
where  $K^*(r)$ is the optimal recovery threshold defined in (\ref{eq:ORT}) and $K_{\textup{PCR}}(r)$ is the recovery threshold of PCR.
\end{theorem}


In Theorem~\ref{thm:newCoding} we focus on the cases of $r \geq 2$. This is because that when $r=1$, the naive scheme where each worker stores and processes a single uncoded sub-matrix trivially achieves the optimal recovery threshold of $n$. To prove Theorem~\ref{thm:newCoding}, we first describe the proposed PCR scheme in the next section, and analyze its recovery threshold $K_{\textup{PCR}}(r)$. Next, in Section~\ref{sec:converse}, we prove a lower bound on the minimum recovery threshold $K^*(r)$, which is no less than half of $K_{\textup{PCR}}(r)$.


\begin{remark} [Comparison with state-of-the-art] As mentioned before, there have been several gradient coding schemes proposed earlier to mitigate stragglers (see e.g., \cite{TLDK-ICML,halbawi2017improving,raviv2017gradient,ye2018communication}), which achieve the recovery threshold of $K_{\textup{GC}}(r)=n-r+1$. Compared with state-of-the-art, the proposed PCR scheme has the following advantages.
\begin{itemize}[leftmargin=*]
\item PCR achieves a much smaller recovery threshold that is within a factor of $2$ from the minimum possible for distributed linear regression. More specifically, for the same computation/storage load $r$, PCR approximately reduces the recovery threshold within each GD iteration by a multiplicative factor of $r/2$. Note that sometimes we want to terminate an iteration when a target number of workers, say $k_{\textup{t}}<n$, have returned their results. This could happen when e.g., we have a large number of data points such that the local data storage/processing becomes the performance bottleneck. In this case, PCR achieves this target recovery threshold with much less storage and computation overhead than the GC schemes. 
\item As we will describe in Section~\ref{sec:PCR}, the key ingredient of PCR is that it directly encodes the data batches stored at the workers rather than coding the results computed from uncoded data as done by GC schemes. To compute the gradient for linear regression, operating directly on the coded data (as done in PCR) gives more useful information to recover the final gradient than combining individual partial gradients computed from uncoded data. This intuitively explains why PCR requires results from much fewer workers. PCR's much smaller recovery threshold significantly reduces both the bandwidth consumption required to deliver the results of the non-straggling workers to the master and the decoding complexity at the master.

\item The state-of-the-art GC schemes are developed for solving general learning problems using distributed gradient descent, where the functions used to compute the gradient from the data and the weight vector can be arbitrary. PCR provides the above advantages by tailoring the data encoding to the algebra underlying the gradient computation of the least-squares regression problems.
\end{itemize}

\end{remark}

\begin{remark}
We note that the proposed PCR scheme is directly applicable for non-linear regression problems using kernel methods. To do that, we simply replace the data matrix $\vct{X}$ with the kernel matrix $\vct{{\cal K}}$, whose entry ${\cal K}_{ij} = k (\vct{x}_i,\vct{x}_j)$ is some kernel function of the data points $\vct{x}_i$ and $\vct{x}_j$. Hence, PCR can be used to optimally combat stragglers for such distributed non-linear regression problems.
\end{remark}

\section{Polynomially coded regression}\label{sec:PCR}
In this section we describe our proposed Polynomially Coded Regression (PCR) scheme, and prove that the recovery threshold reported in Theorem~\ref{thm:newCoding} is achievable. Also, we analyze the computation and communication complexities of PCR, and compare them with those achieved by gradient coding (GC) schemes.

\subsection{Illustrative example}
Before presenting the general PCR scheme, we first use an example to illustrate the key ideas of PCR. In particular, we consider executing a linear regression task on $n=6$ workers, each of whom stores $\frac{r}{n}=\frac{3}{6}=\frac{1}{2}$ fraction of the input feature matrix $\vct{X} = [\vct{X}_0 \; \vct{X}_1 \cdots  \;\vct{X}_5]^\top$.  

\begin{figure}[htbp]
  \centering
  \includegraphics[width=1\textwidth]{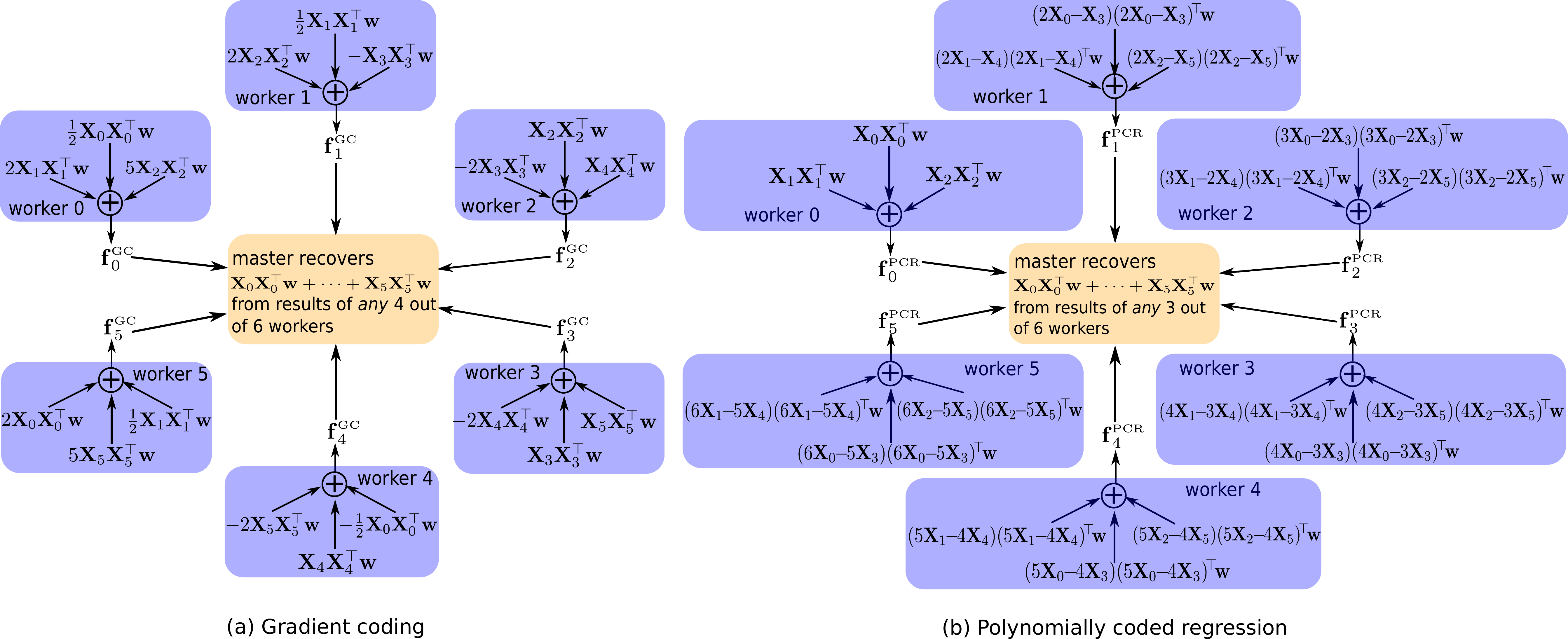}
  \caption{Local computation and communication of the GC scheme and the proposed PCR scheme.}
  \label{fig:example}
\end{figure}

Using the GC scheme\footnote{Here we follow the deterministic construction in~\cite{ye2018communication}.}, each worker~$j$ stores $r=3$ uncoded sub-matrices $\vct{X}_{j}$, $\vct{X}_{j+1}$, and $\vct{X}_{j+2}$. Within each iteration, as shown in Figure~\ref{fig:example}(a), each worker $j$ computes $\vct{X}_i\vct{X}_i^\top\vct{w}$ for each $\vct{X}_i$ stored locally, then creates a linear combination $\vct{f}_j^{\textup{GC}}$ of the 3 local results, and sends it to the master. The master can recover $\sum_{j=0}^5 \vct{X}_j\vct{X}_j^\top\vct{w}$ from the results of any $n-r+1=4$ workers. For example, say workers 1, 2, 3, and 4 finish their computations first, the master can recover the final result as $\sum_{j=0}^5 \vct{X}_j\vct{X}_j^\top\vct{w} = 2\vct{f}_1^{\textup{GC}}-3\vct{f}_2^{\textup{GC}}-3\vct{f}_3^{\textup{GC}}-2\vct{f}_4^{\textup{GC}}$.

For the proposed PCR scheme, in contrast to GC, each worker~$j$ stores 3 \emph{coded} sub-matrices $\tilde{\vct{X}}_{j,0} =(j+1)\vct{X}_0\!-\!j\vct{X}_3$, $\tilde{\vct{X}}_{j,1}=(j+1)\vct{X}_1\!-\!j\vct{X}_4$, and $\tilde{\vct{X}}_{j,2}=(j+1)\vct{X}_2\!-\!j\vct{X}_5$. Within each iteration, as shown in Figure~\ref{fig:example}(b), worker~$j$ simply computes 
\begin{align}
    \vct{f}_j^{\textup{PCR}} = \tilde{\vct{X}}_{j,0}\tilde{\vct{X}}_{j,0}^\top \vct{w} + \tilde{\vct{X}}_{j,1}\tilde{\vct{X}}_{j,1}^\top \vct{w} + \tilde{\vct{X}}_{j,2}\tilde{\vct{X}}_{j,2}^\top \vct{w},
\end{align}
and sends $\vct{f}_j^{\textup{PCR}}$ to the master. We note that by using this data encoding, the local computation at worker~$j$ is essentially evaluating a degree-$2$ polynomial $h(x) = (\vct{X}_0(x+1)\!-\!\vct{X}_3x)(\vct{X}_0^\top \vct{w}(x+1)\!-\!\vct{X}_3^\top\vct{w}x) + (\vct{X}_1(x+1)\!-\!\vct{X}_4x)(\vct{X}_1^\top \vct{w}(x+1)\!-\!\vct{X}_4^\top\vct{w}x) + (\vct{X}_2(x+1)\!-\!\vct{X}_5x)(\vct{X}_2^\top \vct{w}(x+1)\!-\!\vct{X}_5^\top\vct{w}x)$ at the point $x=j$. Therefore, using the computation results from any $3$ of the $6$ workers, the master can interpolate $h(x)$, whose coefficients are functions of the sub-matrices and the weight vector. Having obtained $h(x)$, the master finally evaluates it at $x=0$ and $x=-1$, and sums the results up to find the final result as follows.
\begin{align}
    h(0) \!+\! h(-1) \!=\! \vct{X}_0\vct{X}_0^\top\vct{w} + \vct{X}_1\vct{X}_1^\top\vct{w} + \vct{X}_2\vct{X}_2^\top\vct{w} + \vct{X}_3\vct{X}_3^\top\vct{w} + \vct{X}_4\vct{X}_4^\top\vct{w} + \vct{X}_5\vct{X}_5^\top\vct{w}.
\end{align}

In this example, utilizing data encoding, PCR achieves a recovery threshold of $3$, which improves upon the GC scheme that has a recovery threshold of $4$.

\subsection{Description of PCR and characterization of its recovery threshold}
{\bf Data encoding and placement.} We start by picking $\lceil \frac{n}{r}\rceil$ distinct real numbers $\alpha_0,\alpha_1,\ldots,\alpha_{\lceil \frac{n}{r}\rceil-1}$. Also, we pick $n$ distinct real numbers $\beta_0,\beta_1,\ldots,\beta_{n-1}$, where each $\beta_j$ corresponds to a worker~$j$. For each $j=0,1,\ldots,n-1$, we generate $r$ coded sub-matrices $\{\tilde{\vct{X}}_{j,k}\}_{k=0}^{r-1}$ from the input data matrix $\vct{X} = [\vct{X}_0 \; \vct{X}_1 \cdots  \;\vct{X}_{n-1}]^\top$ as follows, and store them in the local memory of worker~$j$,
\begin{align}
\tilde{\vct{X}}_{j,k} = \sum_{i=0}^{\lceil \frac{n}{r}\rceil-1} \vct{X}_{ri+k} \prod_{i' \neq i}\frac{\beta_j-\alpha_{i'}}{\alpha_i-\alpha_{i'}}, \quad \forall k=0,1,\ldots,r-1.
\end{align}
Here, we note that each coded sub-matrix above is a linear combination of exactly $\lceil \frac{n}{r}\rceil$ uncoded sub-matrices.\footnote{When $\lceil \frac{n}{r}\rceil > \frac{n}{r}$, we append $r\lceil \frac{n}{r}\rceil-n$ zero sub-matrices to the original data matrix $\vct{X}$. That is, for the zero-padded matrix $\vct{X}$, we have $ \vct{X}_n=\vct{X}_{n+1}=\cdots=\vct{X}_{r\lceil \frac{n}{r}\rceil-1} =\vct{0}^{d \times \frac{m}{n}}$.}




{\bf Local computation.} Within each iteration of the distributed GD process, each worker receives the current weight vector $\vct{w}$ from the master. Then, for each $k=0,1,\ldots,r-1$, worker~$j$ multiplies the transpose of the locally stored matrix $\tilde{\vct{X}}_{j,k}$ with $\vct{w}$ to compute an intermediate vector
\begin{align}
        \hat{\vct{f}}_{j,k} = \tilde{\vct{X}}_{j,k}^\top \vct{w}.\label{eq:int1}
\end{align}

Next, worker~$j$ multiplies $\tilde{\vct{X}}_{j,k}$ with the intermediate vector $\hat{\vct{f}}_{j,k}$, for all $k=0,1,\ldots,r-1$, and sums up the results to obtain the final local computation result $\vct{f}_j$ as follows. 
\begin{align}
    \vct{f}_j &= \sum_{k=0}^{r-1}\tilde{\vct{X}}_{j,k}\hat{\vct{f}}_{j,k} = \sum_{k=0}^{r-1}\tilde{\vct{X}}_{j,k}\tilde{\vct{X}}_{j,k}^\top \vct{w}\label{eq:int2}\\
    &=\sum_{k=0}^{r-1}\left(\sum_{i=0}^{\lceil \frac{n}{r}\rceil-1} \vct{X}_{ri+k} \prod_{i' \neq i}\frac{\beta_j-\alpha_{i'}}{\alpha_i-\alpha_{i'}}\right) \left(\sum_{i=0}^{\lceil \frac{n}{r}\rceil-1} \vct{X}_{ri+k}^\top \vct{w} \prod_{i' \neq i}\frac{\beta_j-\alpha_{i'}}{\alpha_i-\alpha_{i'}}\right)
\end{align}

We note that the local computation of worker~$j$ is essentially to evaluate a polynomial $h(x) = \sum_{k=0}^{r-1}\left(\sum_{i=0}^{\lceil \frac{n}{r}\rceil-1} \vct{X}_{ri+k} \prod_{i' \neq i}\frac{x-\alpha_{i'}}{\alpha_i-\alpha_{i'}}\right) \left(\sum_{i=0}^{\lceil \frac{n}{r}\rceil-1} \vct{X}_{ri+k}^\top \vct{w} \prod_{i' \neq i}\frac{x-\alpha_{i'}}{\alpha_i-\alpha_{i'}}\right)$ of degree $2\lceil \frac{n}{r}\rceil-2$ at the point $\beta_j$.

{\bf Aggregation and decoding.} Having received the computation results from the fastest $2\lceil \frac{n}{r}\rceil-1$ workers, whose indices are denoted by $j_0,j_1,\ldots,j_{2\lceil \frac{n}{r}\rceil-2}$, the master uses the evaluations $h(\beta_{j_0}),h(\beta_{j_1}),\ldots,h(\beta_{j_{2\lceil \frac{n}{r}\rceil-2}})$ to interpolate the polynomial $h(x)$, whose coefficients are length-$d$ vectors depending on $\vct{X}$ and $\vct{w}$. Having obtained $h(x)$, the master evaluates it at the points $\alpha_0,\alpha_1,\ldots,\alpha_{\lceil \frac{n}{r}\rceil-1}$. For each $\alpha_i$, we have 
\begin{align}
    h(\alpha_i) &= \sum_{k=0}^{r-1}\left(\sum_{t=0}^{\lceil \frac{n}{r}\rceil-1} \vct{X}_{rt+k} \prod_{t' \neq t}\frac{\alpha_i-\alpha_{t'}}{\alpha_t-\alpha_{t'}}\right) \left(\sum_{t=0}^{\lceil \frac{n}{r}\rceil-1} \vct{X}_{rt+k}^\top \vct{w} \prod_{t' \neq t}\frac{\alpha_i-\alpha_{t'}}{\alpha_t-\alpha_{t'}}\right)\\
    &=\sum_{k=0}^{r-1} \vct{X}_{ri+k}\vct{X}_{ri+k}^\top \vct{w}.
\end{align}

Finally, the master sums up the above evaluation results to obtain the intended computation $\vct{X}^\top\vct{X}\vct{w}$. That is, the master computes
\begin{align}
    \sum_{i=0}^{\lceil \frac{n}{r}\rceil-1} h(\alpha_i) =  \sum_{i=0}^{\lceil \frac{n}{r}\rceil-1} \sum_{k=0}^{r-1} \vct{X}_{ri+k}\vct{X}_{ri+k}^\top \vct{w} = \sum_{j=0}^{n-1} \vct{X}_j\vct{X}_j^\top \vct{w} = \vct{X}^\top \vct{X} \vct{w}.
\end{align}

We thus have demonstrated that the above PCR scheme achieves a recovery threshold of $ K_{\textup{PCR}}(r)= 2\lceil \tfrac{n}{r}\rceil -1$.

\begin{remark}
By choosing $\beta_j = \alpha_j$ for $j=0,1,\ldots,\lceil\frac{n}{r} \rceil-1$, we will have a systematic construction of the PCR scheme. In this case, the local computation results of the first $\lceil\frac{n}{r} \rceil$ workers are the intended polynomial evaluations whose summation is the final required computation for model update.
\end{remark}


\begin{remark}
The above described PCR scheme is motivated by techniques used in a coded computing scheme in~\cite{yu2018straggler}, which is designed to mitigate stragglers in distributed computation of an element-wise vector product. Due to the particular algebraic structure of linear regression, the distributed gradient computation reduces to distributed vector inner product, which can be computed from the element-wise product. While the coding scheme in~\cite{yu2018straggler} is designed for one-shot operation, PCR is used to mitigate the straggler's effect in each iteration of the gradient descent process, which accumulatively provides a substantial reduction on the execution time. Using PCR, coding techniques are directly applied on the input data to further reduce the recovery threshold over gradient coding, which assumes general gradient computation and only codes partial gradients computed from uncoded data.
\end{remark}

\subsection{Complexity analysis of PCR}
In this sub-section we discuss the computation and communication costs of the proposed PCR scheme in each iteration of the GD algorithm, and compare them with those of the GC schemes.

{\bf Computation cost at each worker.} As shown in (\ref{eq:int1}) and (\ref{eq:int2}), for each of the $r$ locally stored coded sub-matrices, each worker first multiplies a $\frac{m}{n} \times d$ matrix with a $d$-dimensional vector, and then multiplies a $d \times \frac{m}{n}$ matrix with a $\frac{m}{n}$-dimensional vector. This incurs a total of $\frac{2rmd}{n}$ element-wise multiplications. On the other hand, for GC schemes, each worker performs the same computations on $r$ uncoded sub-matrices stored locally, incurring the same computation complexity as PCR.

{\bf Communication cost.} The message sent from each worker to the master is a $d$-dimensional real vector, and the master receives $2\lceil \frac{n}{r}\rceil-1$ messages before ending a single iteration. In comparison, for GC, each worker also sends a $d$-dimensional real vector to the master, but now the server needs to wait for the messages from the fastest $n-r+1$ workers.

{\bf Decoding complexity.} The decoding operation at the master is to interpolate $d$ degree-$(2\lceil \frac{n}{r}\rceil-2)$ polynomials, and evaluate each of them at $\lceil \frac{n}{r}\rceil$ points. The complexities of interpolating a degree-$k$ polynomial and evaluating a degree-$k$ polynomial at $k$ points are both $O(k\log^2 k \log \log k)$ (see, e.g.,~\cite{kedlaya2011fast}), hence the decoding complexity at the master of the PCR scheme is $O(d\frac{n}{r}\log^2 \frac{n}{r} \log \log \frac{n}{r})$. When we consider workers with fixed storage size, i.e., $\frac{r}{n}$ is a constant, this decoding complexity does not scale with the number of workers $n$. For the GC schemes, the master decodes the complete gradient by computing a linear combination of the results from the fastest $n-r+1$ workers, each of which is a $d$-dimensional real vector. Therefore, the decoding complexity of GC is $O(d(n-r))$, which scales linearly with the number of workers.\footnote{Here we focus on the case where $d \gg n$, and ignore the computation complexity of finding the coefficients to combine results from non-straggling workers in GC, which is, e.g., quadratic in $(n-r)$ when partial gradients are encoded using Reed-Solomon codes~\cite{halbawi2017improving}.}

To summarize, within each iteration of GD, achieving a much smaller recovery threshold directly helps the PCR scheme to significantly reduce the total amount of communication load from the workers to the master, and the decoding complexity at the master, compared with the GC schemes.

\section{Lower bound on the minimum recovery threshold $K^*(r)$}\label{sec:converse}

Now we prove the converse bound for inequality (\ref{eq:PCR}). To lower bound the recovery threshold $K^*(r)$, we first prove that for any coded computation scheme, the master always needs to wait for at least $\lceil\frac{n}{r}\rceil$ workers to be able to decode the final result. The factor of $2$ characterization in Theorem \ref{thm:newCoding} directly follows from this lower bound, as $\tfrac{1}{2}K_{\textup{PCR}}(r)<\lceil\frac{n}{r}\rceil$.

To prove the latter statement, we aim to show that, for any coded computation scheme and any subset $\mathcal{N}$ of workers, if the master can recover $\vct{X}^\top\vct{X}\vct{w}$ given the results from workers in $\mathcal{N}$, then we must have $|\mathcal{N}|\geq \lceil\frac{n}{r}\rceil$.
Suppose the condition in the above statement holds, then we can find encoding, computation, and decoding functions such that for any possible values of $\vct{X}$ and $\vct{w}$, the composition of these functions returns the correct output.


Note that within a GD iteration, 
each worker performs its local computation only based on its locally stored coded sub-matrices and the weight vector $\vct{w}$. Hence, if the master can decode the final output from the results of the workers in a subset $\mathcal{N}$, then the composition of the decoding function and the computing functions of these workers essentially computes $\vct{X}^\top\vct{X}\vct{w}$ only using the coded sub-matrices stored by these workers and the vector $\vct{w}$. Hence, if any class of input values $\vct{X}$ gives the same coded sub-matrices for each worker in $\mathcal{N}$, then the product $\vct{X}^\top\vct{X}\vct{w}$ must also be the same given any $\vct{w}$. 


Now we consider the class of input matrices $\vct{X}$ such that all coded sub-matrices stored at workers in $\mathcal{N}$ equal the values of the corresponding coded sub-matrices when $\vct{X}$ is zero. 
Since $\vct{0}^\top\vct{0}\vct{w}$ is zero for any $\vct{w}$, 
$\vct{X}^\top\vct{X}\vct{w}$ must also be zero for all matrices $\vct{X}$ in this class and any $\vct{w}$. However, for real matrices $\vct{X}=\vct{0}$ is the only solution to that condition. Thus, zero matrix must be the only input matrix that belongs to this class.


Recall that all the encoding functions are assumed to be linear. We consider the collection of all encoding functions that are used by workers in $\mathcal{N}$, which is also a linear map. As we have just proved, the kernel of this linear map is $\{\vct{0}\}$. Hence, its rank must be at least the dimension of the input matrix, which is $dm$.
On the other hand, its rank is upper bounded by the dimension of the output, where each encoding function from a worker contributes at most 
$\frac{rdm}{n}$. Consequently, the number of workers in $\mathcal{N}$ must be at least $\lceil\frac{n}{r}\rceil$ to provide sufficient rank to support the computation. This concludes the converse proof. 



\section{Experiments}
In this section, we present the experimental results on Amazon EC2 clusters. In particular, we empirically compare the performance of our proposed PCR scheme with the GC scheme in~\cite{TLDK-ICML} (specifically, the cyclic repetition scheme), the naive uncoded scheme for which there is no data redundancy among the workers, and the batched coupon's collector (BCC) scheme in~\cite{li2018near} that randomly places uncoded data batches onto workers.

\subsection{Experimental Setup}
We trained a linear regression model distributedly using Nesterov's accelerated gradient descent. We ran the master and all workers over \texttt{t2.micro} instances. We implemented all different schemes in python, and used MPI4py~\cite{dalcin2011parallel} for message passing between instances. Before the GD iterations, each worker stores a certain amount of data in its local memory. In the $t$th iteration, having received the latest model $w^{(t)}$ from the master, each worker computes the local result using the stored data and $w^{(t)}$, and sends it back to the master asynchronously using \texttt{Isend()}. As soon as the master gathers enough results from the workers, it computes the final gradient and updates the model. 



\noindent {\bf Data Generation.} 
We create a synthetic training dataset $D$ consisting of $m$ input-output pairs, i.e., $D=\{(\vct{x}_1,y_1),(\vct{x}_2,y_2),\ldots,(\vct{x}_m,y_m)\}$. To do that, we first generate a true weight vector $\vct{w}^{*}$ whose components are i.i.d. and uniformly sampled from $[0,1]$. Then, we generate each input vector $\vct{x}_i$ with $d=7000$ features according to a normal mixture distribution $\frac{1}{2}\times \mathcal{N}(\vct{\mu}_{1},\vct{I})+\frac{1}{2}\times \mathcal{N}(\vct{\mu}_{2},\vct{I})$, where $\vct{\mu}_{1}=\frac{1.5}{d}\vct{w}^{*}$ and $\vct{\mu}_{2}=\frac{-1.5}{d}\vct{w}^{*}$, and its corresponding output label $y_i=\vct{x}_i^\top\vct{w}^{*}$.

We run GD for 100 iterations in various settings with different sizes of the dataset and numbers of workers. To better compare the effectiveness of straggler mitigation using different schemes, in some scenarios we also artificially introduced some delay to the worker instances (using \texttt{time.sleep()}). In particular, in each iteration, we imposed a $0.5$ seconds delay on each worker with probability $5\%$. We summarize the four considered experiment scenarios in the following Table~\ref{table:scenarios}.

\begin{table}[htbp]
  \centering
  \begin{tabular}{| c | c | c | c | }
    \hline
    scenario index & \# of data points ($m$)& \# of workers ($n$) &artificial stragglers  \\ \hline
    1 & 8000  &40  & No\\ \hline
    2 & 8000  &40  & Yes\\ \hline
    3 & 6000  &30  & No\\ \hline
    4 & 6000  &30  & Yes\\ \hline
  \end{tabular}
  \vspace{1mm}
  \caption{Experiment scenarios.}
\label{table:scenarios}
\end{table}





\subsection{Results}
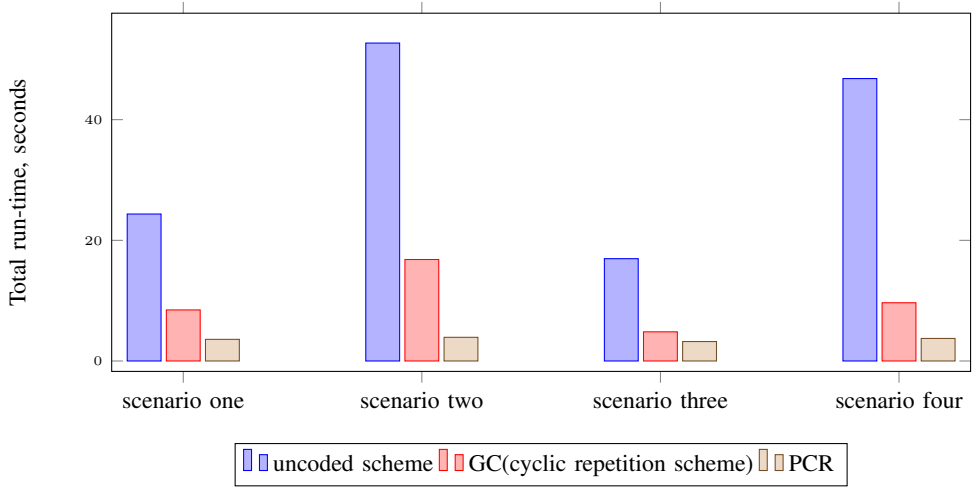
\begin{figure}[htbp]
\centering
\begin{tikzpicture}{center} \begin{axis}[
    ybar,
    yticklabel style = {font=\tiny},
    bar width=.45cm,
    width=13cm,
    height=.35\textwidth,
    legend style={at={(0.5,-0.2)},
      anchor=north,legend columns=-1},
    ylabel={Total run-time, seconds},style = {font=\small},
    xticklabels={scenario one,scenario two,scenario three,scenario four}, style = {font=\small},
    xtick=data,
    ]
\addplot coordinates {(1,24.362) (1.5,52.700)(2,16.961)(2.5,46.817)};
\addplot coordinates {(1,8.464) (1.5,16.821)(2,4.833)(2.5,9.653)};
\addplot coordinates {(1,3.587) (1.5,3.925)(2,3.221)(2.5,3.736)};
\legend{uncoded scheme,GC(cyclic repetition scheme),PCR}
\end{axis}
\end{tikzpicture}
\caption{Run-time comparison of the uncoded, the GC (cyclic repetition), and the PCR schemes on EC2 clusters. Scenarios 1 and 2 use $n=40$ workers, while scenarios 3 and 4 use $n=30$ workers. In scenarios 2 and 4, we artificially delayed each worker for $0.5$ s with probability $5\%$. In all scenarios, the computation/storage load $r$ of GC and PCR were chosen to minimize their total run-times.}
\label{fig:run-time}
\end{figure}

\begin{figure}[!htbp]
  \centering
  \includegraphics[width=0.9\textwidth]{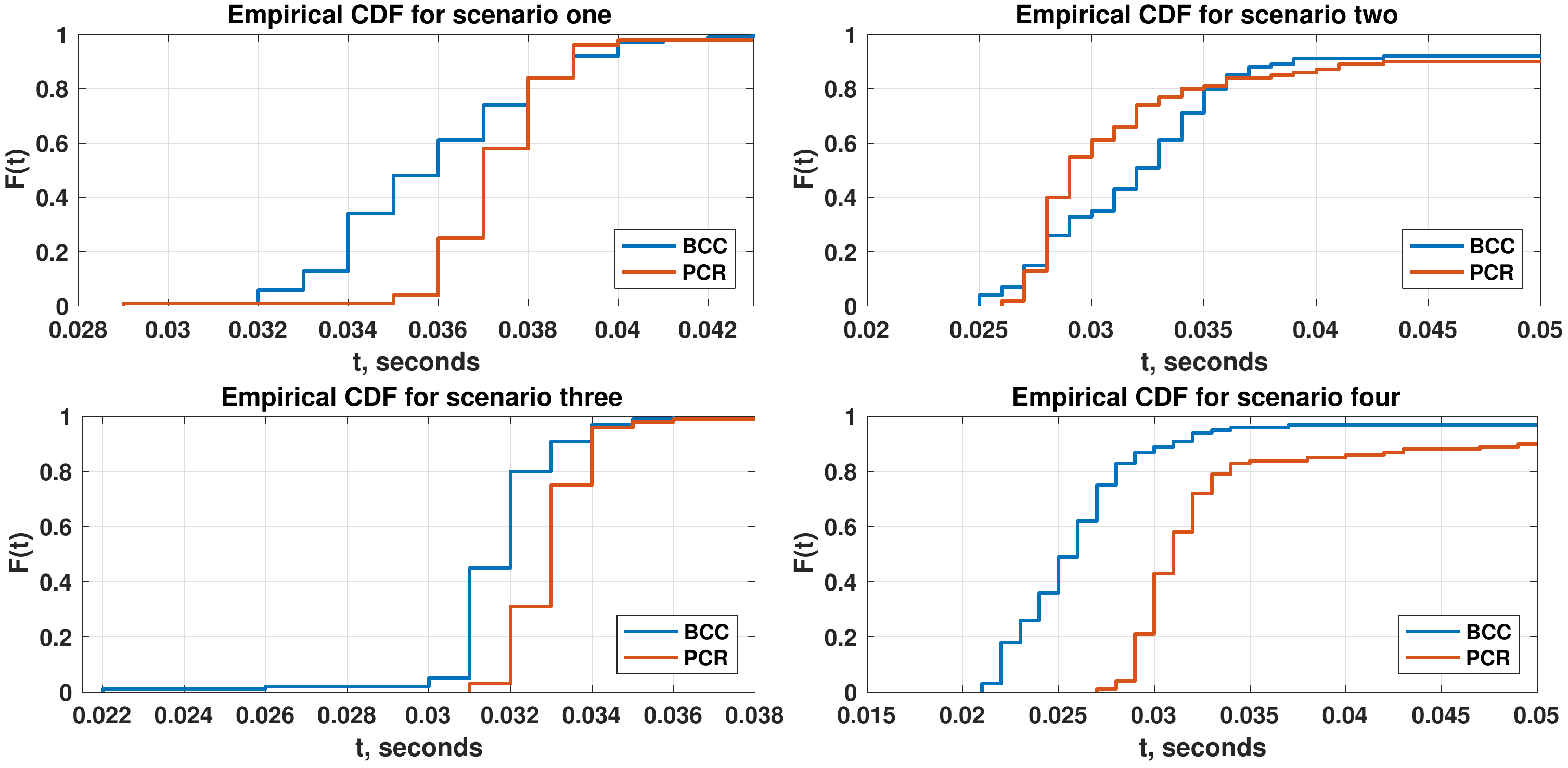}
  \caption{Empirical CDF of run-time per iteration of PCR and BCC schemes in four scenarios.}
  \label{fig:cdf}
\end{figure}

For the uncoded scheme, each worker stores and processes
$r=1$ data batch. For the GC, PCR, and BCC schemes, we selected the optimal computation/storage load $r$ subject to the memory size of the \texttt{t2.micro} instance to minimize the total run-time. For the uncoded, GC, and PCR schemes that have fixed recovery thresholds across iterations, we plot their run-times in all four experiment scenarios in Figure~\ref{fig:run-time}, and also list the breakdowns of their run-times in Tables~\ref{table:scenario one} to \ref{table:scenario four}. The computation time was measured as the summation of the maximum local processing time among all non-straggling workers, over 100 iterations. The communication time is computed as the difference between the total run-time and the computation time. Finally, we plot the CDFs of the per iteration run-time for the PCR and BCC schemes in the four scenarios in Figure~\ref{fig:cdf}. 

\begin{table}[!htbp]
  \centering
  \begin{tabular}{|c | c | c| c | c | c |}
    \hline
     \multirow{2}{*}{schemes} & \# batches processed & recovery & communication  &computation  & \multirow{2}{*}{total run-time} \\
     & at each worker ($r$) & threshold & time & time&\\
     \hline
    uncoded & 1 & 40   &24.125 s &0.237 s &  24.362 s\\ \hline
    GC & 10 &31&6.033 s  & 2.431 s  &8.464 s\\ \hline
    PCR & 10 & 7&1.719 s & 1.868 s&  3.587 s\\ \hline
  \end{tabular}
  \vspace{1mm}
  \caption{Breakdowns of the run-times in scenario one with $n=40$ workers.}
\label{table:scenario one}
\end{table}

\begin{table}[!htbp]
  \centering
  \begin{tabular}{|c | c | c| c | c | c |}
    \hline
     \multirow{2}{*}{schemes} & \# batches processed & recovery & communication  &computation  & \multirow{2}{*}{total run-time} \\
     & at each worker ($r$) & threshold & time & time&\\
     \hline
    uncoded & 1 & 40 & 7.928 s & 44.772 s & 52.700 s\\ \hline
    GC & 10 & 31 & 14.42 s  & 2.401 s  & 16.821 s\\ \hline
    PCR & 10 & 7 & 2.019 s & 1.906 s &  3.925 s\\ \hline
  \end{tabular}
  \vspace{1mm}
  \caption{Breakdowns of the run-times in scenario two with $n=40$ workers.}
\label{table:scenario two}
\end{table}



\begin{table}[!htbp]
  \centering
  \begin{tabular}{|c | c | c| c | c | c |}
    \hline
     \multirow{2}{*}{schemes} & \# batches processed & recovery & communication  &computation  & \multirow{2}{*}{total run-time} \\
     & at each worker ($r$) & threshold & time & time&\\
     \hline
    uncoded & 1 & 30  & 16.731 s & 0.230 s & 16.961 s\\ \hline
    GC & 10 & 21 & 2.617 s  & 2.216 s & 4.833 s\\ \hline
    PCR & 10 & 5 & 1.373 s & 1.848 s &  3.221 s\\ \hline
  \end{tabular}
  \vspace{1mm}
  \caption{Breakdowns of the run-times in scenario three with $n=30$ workers.}
\label{table:scenario three}
\end{table}


\begin{table}[!htbp]
  \centering
  \begin{tabular}{|c | c | c| c | c | c |}
    \hline
     \multirow{2}{*}{schemes} & \# batches processed & recovery & communication  &computation  & \multirow{2}{*}{total run-time} \\
     & at each worker ($r$) & threshold & time & time&\\
     \hline
    uncoded & 1 & 30 & 9.556 s & 37.261 s & 46.817 s\\ \hline
    GC & 10 & 21 & 7.393 s & 2.260 s  & 9.653 s\\ \hline
    PCR & 10 & 5 & 1.857 s & 1.879 s&  3.736 s\\ \hline
  \end{tabular}
  \vspace{1mm}
  \caption{Breakdowns of the run-times in scenario four with $n=30$ workers.}
\label{table:scenario four}
\end{table}

Based on the results, we draw the following conclusions.
\begin{itemize}[leftmargin=*]

\item In scenarios one and two, PCR allows the system to tolerate $24$ more stragglers than GC. In scenarios three and four, PCR tolerates 16 more stragglers than GC. Over the four scenarios, with naturally occurring stragglers, PCR speeds up the job execution of the uncoded scheme by $5.27\times \!\sim\! 6.79\times$, and the GC scheme by $1.50\times \!\sim\! 2.36\times$. With artificial stragglers, PCR speeds up the job execution of the uncoded scheme by $12.53\times \!\sim\! 13.43\times$, and the GC scheme by $2.58\times \!\sim\! 4.29\times$.

\item The run-time improvements of PCR over the uncoded and the GC schemes become more significant as the size of the network increases. This is because that the recovery threshold of PCR scales inversely proportionally to the computation/storage load at each worker, and it increases much slower than the other two schemes when increasing the number of workers. Due to the same reason, we would also expect a much more significant execution speedup from PCR when more computation/storage is available at each worker.



\item In all four experiment scenarios, PCR achieves both a smaller computation time and a smaller communication time than GC. Given that PCR achieves a much smaller recovery threshold, much fewer workers are required to finish their local computations, resulting in a shorter computation time. On the other hand, since the computation results of the workers have the same size, a smaller recovery threshold directly leads to a smaller communication load to the master.

\item With the same optimal computation/storage load ($r=10$) as PCR, the BCC scheme achieves an average recovery threshold that is close to the fixed recovery threshold of PCR in each scenario. However, since BCC randomly places the data batches onto workers and the resulting recovery threshold varies across iterations, we can see from Figure~\ref{fig:cdf} that PCR exhibits a more concentrated and predictable run-time in each GD iteration than BCC.



\end{itemize}

\bibliographystyle{IEEEtran}
\bibliography{reference}

\end{document}